# Giant magneto-cubic in-plane Hall effect in a nonmagnetic material

Jie Chen[1,8], Jin Cao[2,8], Yue Lu[1], Hang Li[1], Xiaodong Zhou[1], Xuekui Xi[3], Orest Pavlosiuk[4], Piotr Wiśniewski[4], Dariusz Kaczorowski[4], Yong-Chang Lau[3,5]*, Cong Xiao[6]*, Yue Li[1], Yong Jiang[1], Wenhong Wang[1]* and Shengyuan A. Yang[7]

[1]School of Electronic and Information Engineering, Tiangong University, Tianjin 300387, China

[2]Institute of Applied Physics and Materials Engineering, University of Macau, Macau 999078, China

[3]Beijing National Laboratory for Condensed Matter Physics, Institute of Physics, Chinese Academy of Sciences, Beijing 100190, China

[4]Institute of Low Temperature and Structure Research, Polish Academy of Sciences, Okólna 2, Wrocław 50-422, Poland

[5]University of Chinese Academy of Sciences, Beijing 100049, China

[6]Interdisciplinary Center for Theoretical Physics and Information Sciences (ICTPIS), Fudan University, Shanghai 200433, China

[7]Research Laboratory for Quantum Materials, Department of Applied Physics, The Hong Kong Polytechnic University, Kowloon, Hong Kong SAR 999077, China

[8]These authors contributed equally: J. Chen, and J. Cao

*Corresponding author E-mail: wenhongwang@tiangong.edu.cn; congxiao@fudan.edu.cn；yongchang.lau@iphy.ac.cn

**Abstract**

In-plane Hall effect (IPHE) triggered by an external magnetic field applied in the transport plane has attracted significant experimental attentions in recent few years [1-6]. However, most experiments focus on magnetic materials, where the existence of magnetic ordering may complicate understanding the physics behind, and the relatively small signal magnitudes limit the application of the effect. Here, we report a giant IPHE in a nonmagnetic half-Heusler compound LuAuSn, with a magnitude exceeding all the previously reported values. A $2\pi/3$-period of IPHE and the consistent cubic dependence on the magnetic field are observed, realizing the long-sought theoretical prediction of magneto-cubic IPHE under threefold rotational symmetry[7-9] in an unexpected material. The scaling law analysis and first-principles calculations indicate that extrinsic side jump and skew scattering processes from both impurity and phonon scatterings dominate the observed effect. These findings unravel a new type of magneto-nonlinear IPHE, and its large magnitude and wide-temperature operation may open the door to practical applications of IPHE.

Hall effect is one of the most important physical phenomena in condensed matter physics [10]. Experimentally, the Hall effect is most usually manifested with the external magnetic field applied along the normal direction of the transport plane defined by the driving electric current and the generated transverse voltage (Fig. 1**a**). In 2018, the observation of Hall effect in $ZrTe_5$ induced by magnetic field applied within the transport plane unravels the in-plane Hall effect (IPHE) (Fig.1**b**) [7, 11], which was later systematically studied in VS-$VS_2$, showing a perfect linear dependence on the in-plane magnetic field and a $2\pi$-periodic angular dependence [3]. Such studies recall earlier theoretical investigations on the quantized anomalous Hall effect induced by in-plane magnetization [7, 12, 13], and then inspired intensive experimental attentions on IPHE in magnetic materials, such as $Fe_3Sn_2$, $RuO_2$, CuPt/CoPt heterostructure, $EuCd_2Sb_2$, Fe films, $Co_3Sn_2S_2$, and $EuZn_2Sb_2$ [5, 6, 11, 14-18]. The key issues in these studies are how does the magnetic order parameter respond to the magnetic field and how does this interplay result in the observed angular dependence and scaling on the field [5, 19]. On the other hand, theoretical studies of IPHE in recent few years mostly focused on nonmagnetic materials [8, 9, 13, 20-28]. This is because nonmagnetic materials avoid the complexity introduced by the intricate interplay of magnetic field and the magnetic order parameter. Therefore, to observe this fundamental transport effect in nonmagnetic materials is critical for revealing the IPHE physics induced by magnetic field alone. However, compared to the large body of IPHE experiments in magnetic materials, IPHE in nonmagnetic materials has so far only been achieved in $ZrTe_5$ [1, 29, 30], $LaAlO_3$/$SrTiO_3$ interfaces [4], and $IrO_2$ [14]. While the IPHE in $IrO_2$ has been attributed to Lorentz force, the origins of those in $ZrTe_5$ and $LaAlO_3$/$SrTiO_3$ interfaces are still elusive [1, 4, 29, 30], partly due to their complex dependences on magnetic field beyond simple low order power-law dependence. None of the theoretically proposed anomalous mechanisms for IPHE, including the field induced Berry curvature [9, 20, 22, 25, 26], side jump and skew scattering [9, 31], has been firmly detected in experiments.

In this work, we make a step forward by observing the IPHE in a nonmagnetic half-Heusler compound LuAuSn, which shows a $2\pi/3$-period angular dependence on the magnetic field orientation and a cubic scaling on the field strength up to $B = 3$ T, realizing the long-sought theoretical prediction of magneto-cubic IPHE under $C_{3z}$ symmetry in an unexpected material [7-9]. The scaling law analysis and the comparison with the results of first-principles calculations indicate that extrinsic side jump and skew scattering processes from both impurity and phonon scatterings dominate the observed IPHE. At $B = 3$ T and $T = 2$ K, the Hall conductivity is 306 $\Omega^{-1}cm^{-1}$, which is more than one order of magnitude larger than the value in ZrTe$_5$ measured at similar temperature and field and is also larger than all the up-to-now reported in-plane Hall conductivities in magnetic materials. The IPHE conductivity decreases rapidly with increasing temperature but still remains sizable up to room temperature. The nonmagnetic nature of the material, the clean $B^3$ scaling at a wide range of magnetic field, the giant in-plane Hall conductivity, and the broad operating temperature range jointly make LuAuSn an ideal platform for investigating IPHE and related in-plane Hall families.

**Results**

**Cubic-*B* dependent IPHE in C$_{3z}$-symmetric systems**

We first review the IPHE from a symmetry perspective. As a time-reversal-odd effect, Hall effects require the presence of time-reversal-breaking fields, such as the magnetic field in nonmagnetic materials. The Hall effect can be described by a magnetic-field-dependent Hall vector $\boldsymbol{E}^H = \boldsymbol{\rho}^H(\boldsymbol{B}) \times \boldsymbol{j}$, where $\boldsymbol{\rho}^H = \left(\rho^H_{yz}, \rho^H_{zx}, \rho^H_{xy}\right)$, $\boldsymbol{j}$ is the applied current density, and $\boldsymbol{E}^H$ is the induced Hall electric field, which leads to the experimentally measured Hall voltage. The Hall vector is an odd function of magnetic field: $\boldsymbol{\rho}^H(\boldsymbol{B}) = -\boldsymbol{\rho}^H(-\boldsymbol{B})$. The component of the Hall vector perpendicular to the magnetic field applied in the transport plane corresponds to the IPHE of interest. Importantly, the IPHE is independent of the relative direction between the magnetic field and the applied current density [12]. This distinguishes it from the planar Hall effect, which is essentially an anisotropic magnetoresistance effect. Since inversion symmetry imposes no constraints on the Hall vector, it is sufficient to consider the constraints imposed by proper rotational symmetries. Consider an infinitesimal rotation described by $\mathfrak{R} = 1 - d\theta \boldsymbol{L} \cdot \boldsymbol{n}$, where $\boldsymbol{n}$ is the direction of the rotation axis, and $L_i^{jk} = \epsilon_{ijk}$ are the generators of $\mathfrak{so}(3)$ Lie group with $\epsilon_{ijk}$ being the Levi-Civita symbol. This infinitesimal rotation enforces $\mathfrak{R}\boldsymbol{\rho}^H(\mathfrak{R}^{-1}\boldsymbol{B}) = \boldsymbol{\rho}^H(\boldsymbol{B})$, which leads to a differential equation: $\boldsymbol{n} \times \boldsymbol{\rho}^H = (\boldsymbol{n} \times \boldsymbol{B}) \cdot \boldsymbol{\nabla}_B \boldsymbol{\rho}^H$. To fulfill this condition, the Hall vector must be parallel to the magnetic field. This implies that the out-of-plane Hall effect is unrestricted by rotational symmetry, while the IPHE requires reducing symmetry.

Regarding the discrete crystal symmetries, it is sufficient to analyze the Laue point group symmetries. Consider a magnetic field in $xy$ plane and a Hall vector along $z$ direction, under $C_{2z}$ symmetry, $\rho_{xy}^H$ remains unchanged, while the in-plane magnetic field reverses sign. Therefore, the IPHE is forbidden. Due to the same reason, the IPHE is also forbidden under $C_{4z}$, and $C_{6z}$ symmetries. On the other hand, the $C_{3z}$-symmetric may allow IPHE. It should be noted $C_{3z}$ forbids IPHE in the linear order of $B$[12], and the leading order effect should exhibit $B^3$ dependence (Figs.1**d** and 1**f**). We note among the fourteen Laue point groups, $\bar{1}$, $2/m$, $4/m$, $\bar{3}$, and $6/m$ allow $B^1$-dependent IPHE (Figs.1**c** and 1**e**). The $\bar{3}m$ Laue group rules out $B$-linear IPHE but permits a $B^3$ IPHE. For the remaining three Laue point groups, only the out-of-plane Hall effect is allowed. A $B^3$-dependent IPHE has recently been observed in a magnetic Weyl semimetal $Co_3Sn_2S_2$. However, the $B^3$-dependent IPHE is in principle allowed in nonmagnetic materials without the impact of magnetic ordering, though it has not yet been experimentally demonstrated.

LuAuSn is a half-Heusler compound with a non-centrosymmetric crystal structure, which belongs to $\bar{4}3m$ point group (Fig. 1**g**) [32, 33]. To observe the IPHE, we synthesized high quality LuAuSn single crystal along (111) plane with Sn-flux method

[34]. The (111) plane is natural growth surfaces and can easily obtain the flat plane that can reduce the misleading in IPHE measurement. The (111) plane of single crystal are determined by X-ray diffraction and Laue diffraction (Fig. S1). As shown in Fig. 1**g**, the (111) plane is represented by the gray plane, where the Lu, Sn, Au atomic layers are stacked in order along [111] direction. Fig. 1**h** shows the top view of (111) plane. It exhibits $3m$ point group symmetry, consisting of $C_{3z}$ rotation and three mirror symmetries as highlighted by dashed lines. It meets the mentioned symmetry requirement for observing the $B^3$ IPHE. Due to the $C_{3z}$ symmetry, the IPHE will show $2\pi/3$ period angular dependence by rotating the magnetic field (Fig. 1**j**), which is distinguished from the $\pi$-period planar Hall response.

**Observation in nonmagnetic LuAuSn**

The temperature dependent resistivity $\rho_{xx}$ shows typical metallic behavior at zero magnetic field for sample #1 (see Fig. S2**a**). We first investigate the conventional out-of-plane Hall effect, which as mentioned exists even in the isotropic systems. The magnetic field was applied along the normal direction of the (111) plane, and the current was applied along the $[\bar{1}12]$ direction. The Hall resistivity was identified as the field-odd part of the measured transverse resistivity: $\rho_{xy}^H(\boldsymbol{B}) = [\rho_{xy}(\boldsymbol{B}) - \rho_{xy}(-\boldsymbol{B})]/2$. As shown in the inset of Fig. S2**c**, $\rho_{xy}^H$ exhibits linear dependence on the magnetic field across temperature range from 2 K to 300 K and $B \leq 3T$. For fixed $B = 3\,T$, we rotated the magnetic field in the $zy$ plane (see Fig. 2**a**), and the measured Hall resistivity is given in Fig. 2**b**. The Hall signal persists even when the magnetic field lies in the plane, and it is even more pronounced than in the out-of-plane configuration. Fig.2**d** shows the Hall resistivity measured by rotating the in-plane magnetic field, where $\varphi$ is defined as the angle relative to the $[1\bar{1}0]$ axis (Fig. 2**c**). The angular dependence can be well decomposed as:

$$\rho_{xy}^H = \rho_{xy}^i \sin(3\varphi) + \rho_{xy}^o \sin(\varphi + \varphi_0) \qquad (1)$$

and the decomposed components are plotted in Fig. 2**e**. Here, the first term is identified as the in-plane Hall resistivity, and the second term arises from a small misalignment of the magnetic field, with $\varphi_0$ quantifying the deviation between the (111) plane and the magnetic field rotation plane. This decomposition is reasonable given the small out-of-plane component of $B$ field. The $2\pi/3$ period angular dependence is consistent with the IPHE in a $C_{3z}$-symmetric system. Moreover, the in-plane component shows $\sin(3\varphi)$ dependence with no phase shift. This can be explained by the presence of $\mathcal{M}_y$ symmetry, since it imposes the condition $\rho_{xy}^i(90°-\varphi) = -\rho_{xy}^i(90°+\varphi)$. The extracted in-plane Hall resistivity curves at different magnetic fields are shown in Fig. 2**f**. All of them exhibit a $\sin(3\varphi)$ angular dependence. As shown in Fig. 2**g**, $\rho_{xy}^i$ shows a good agreement with cubic $B$-dependence fitting below 3 T. This for the first time verifies the magnetic-cubic IPHE in nonmagnetic compound.

A key distinction between the IPHE and the planar Hall effect is that the IPHE is independent of the relative direction between the magnetic field and the applied current [12]. To verify this in LuAuSn, we designed an experiment in which the magnetic field direction is fixed relative to the crystal, while the driving current is applied along two inequivalent crystallographic directions, the $[1\bar{1}0]$ axis (Fig.3**a**) and its perpendicular direction, the $[\bar{1}12]$ axis (Fig.3**d**). Figs.3**b** and 3**e** depict the longitudinal resistivity ($\rho_{xx}$ and $\rho_{yy}$), both of which show a $\pi$ period oscillation. It is independent of the relative orientation between the magnetic field and the crystal axis. Both configurations show maximal resistivity when $B \perp I$. The two resistivities $\rho_{xx}$ and $\rho_{yy}$ differ by approximately a $\pi/2$ phase, consistent with the $3m$ point group symmetry. Fig. 3**c** and 3**f** present the IPHE resistivity measured in the two configurations. They show identical results with no phase difference. This again confirms the IPHE nature of our observation, which is distinguished from an in-plane AMR effect.

**Giant in-plane Hall conductivity**

We next extract the in-plane Hall conductivity and investigate its temperature and magnetic field dependence. The conductivity were obtained by $\sigma_{xx} = \rho_{yy}/(\rho_{xx}\rho_{yy} - \rho_{xy}\rho_{yx})$ and $\sigma_{xy}^H = -\rho_{xy}^H/(\rho_{xx}\rho_{yy} - \rho_{xy}\rho_{yx})$, where $\rho_{xy}\rho_{yx} = \rho_{(xy)}^2 - (\rho_{xy}^H)^2$, with $\rho_{(xy)}$ being the field-even part of the measured transverse resistivity $\rho_{(xy)}(B) = [\rho_{xy}(B) + \rho_{xy}(-B)]/2$. Then, the in-plane Hall conductivity

$\sigma_{xy}^i$ is identified as the $2\pi/3$ period component. Figs.4**a** and 4**b** show the angular-dependence of the anisotropic longitudinal conductivity $\sigma_{xx}$ under an in-plane magnetic field, and $\sigma_{xy}^i$ at 2 K- 300 K. The magnetic field magnitude was fixed at 3 T. The maximum value of $\sigma_{xy}^i$ reaches up to 306 S/cm at 2K. Fig.4**d** presents the temperature dependence of $\sigma_{xy}^i$ under magnetic fields of 3 T, 6 T and 9 T, respectively. The magnetic field was fixed along the $[\bar{1}12]$ direction ($\varphi = \pi/2$), where the IPHE is the most significant due to the $\sin(3\varphi)$ factor. The in-plane Hall conductivity decreases with increasing temperature but remains finite even at room temperature (17 S/cm at 9 T, 6 S/cm at 6T, and 1 S/cm at 3T). The insert figure shows the corresponding in-plane Hall angle (IPHA). It can reach about 0.1, comparable to conventional anomalous Hall effect in topological semimetals [35-37]. Fig.4**e** also shows the magnetic field dependence of $\sigma_{xy}^i$ at 2 K. Below 3 T, the IPHE follows a clear cubic dependence. It gradually deviates from a cubic dependence at higher magnetic fields. The out-of-plane Hall conductivity $\sigma_{xy}^o$ at 2 K is also presented. It shows a two-carrier type normal Hall curves, which is consistent with the band structure, the hole and electron pockets at X and Γ points [38]. The different behaviors of $\sigma_{xy}^i$ and $\sigma_{xy}^o$ hint fundamentally different physical origins. Notably, the IPHE conductivity of LuAuSn at 2 K and 3 T is an order of magnitude larger than that of non-magnetic ZrTe$_5$. It is even larger than all magnetic IPHE systems (Fig.4**f**). Moreover, the in-plane Hall conductivity in LuAuSn can be further enhanced to approximately 687 $\Omega^{-1}cm^{-1}$ at 7 T.

**Discussion on the physical mechanisms**

To gain more insight into the observed in-plane Hall effect, we performed a scaling relation analysis between the in-plane Hall conductivity and the longitudinal conductivity. Considering two sources of scattering from impurities and phonons, the scaling relation is expressed as [39]:

$$\sigma_{xy}^i = C^{in} + C_4 + (C_3 - 2C_4)\sigma_{xx0}^{-1}\sigma_{xx} + (C_2 + C_4 - C_3 + C_1\sigma_{xx0} + D\sigma_{xx0}^2)\sigma_{xx0}^{-2}\sigma_{xx}^2 \quad ,$$

(2)

where $\sigma_{xx0}$ is the residual conductivity measured at $2\,K$. The fitting parameters are defined as follows: $C_1 = c_0^{csk}$, $C_2 = c_0^{sj} + c_{00}^{isk}$, $C_3 = c_0^{sj} + c_1^{sj} + c_{01}^{csk}$, and $C_4 = c_1^{sj} + c_{11}^{isk}$. Here, $C^{in}$ denotes the intrinsic contribution. The superscripts $sj$, $csk$ and $isk$ refer to contributions involving side-jump, conventional skew scattering, and intrinsic skew scattering, respectively [40]. The subscripts 0 and 1 in these $c$'s denote scattering with impurities and phonons, respectively. $C_1$ and $C_2$ include only impurity scatterings, $C_4$ includes only phonon scattering, whereas $C_3$ includes both impurity and phonon scatterings as well as their compositions. An additional term $D$ is included to represent the in-plane Hall response due to the Lorentz force.

We first determined the intrinsic and Lorentz force contributions from first-principles calculations (see details in Methods). As shown in Fig. **5b**, under an in-plane magnetic field of $3\,T$ applied along $[\bar{1}\bar{1}2]$ direction, the intrinsic Hall conductivity is about 1 S/cm near the Fermi level, which is two orders of magnitude smaller than the experimentally observed value. The Lorentz force contribution is also negligibly small (Fig. **5c**). These results suggest that the observed IPHE most likely originates from side-jump and skew scattering mechanisms, and the $C^{in}$ and $D$ terms in the scaling relation can be neglected. The in-plane Hall conductivity as a function of $\sigma_{xx0}^{-1}\sigma_{xx}$ is shown in Fig. **5d** with $\sigma_{xx}$ in the temperature range from $20\,K$ to $125\,K$, which fits well with Eq. (2). Each term in Eq. (2) is shown in the inset of Fig. **5d**, revealing that all terms are equally important. The fitting result is given by $C_2 + C_1\sigma_{xx0} = 311\,\Omega^{-1}cm^{-1}$, $C_3 = -855\,\Omega^{-1}cm^{-1}$, and $C_4 = 664\,\Omega^{-1}cm^{-1}$. It indicates that both impurity and phonon scatterings play equally important roles. In this case, the scaling relation cannot be further simplified, and separating the side-jump and skew-scattering contributions is in principle impossible in scaling analysis[39]. We note the residual resistance ratio of our sample is 3 (Fig. **S2a**), which may account for the considerable $C_2 + C_1\sigma_{xx0}$ contribution.

**Conclusion and outlook**

We have identified the IPHE in a new material -- half-Heusler compound LuAuSn,

which is not only the first observation of magneto-cubic IPHE in a three-dimensional nonmagnetic material but also paves the way to room-temperature researches and device applications of IPHE. The nonmagnetic material platform LuAuSn for IPHE is free from the limitations of magnetic materials in generally low operating temperatures and intricate interplay of magnetic ordering and external fields, thus rendering an ideal playground for understanding both fundamental and applicational aspects of IPHE physics. This study may also stimulate more future efforts to search in nonmagnetic materials the room-temperature IPHE, its thermoelectric and thermal analogs driven by temperature gradient, i.e., in-plane Nernst effect and in-plane thermal Hall effect, as well as their counterparts in nonlinear order of electric current [41] and temperature gradient. Such in-plane Hall families in nonmagnetic materials may constitute an attractive part of transport studies in the near future.

The revealed physical mechanisms are different from previous IPHE experiments which are mainly attributed to the intrinsic Berry-phase effects [3, 4] or classical Lorentz force [11, 14, 18], unraveling the important role of side jump and skew scattering in IPHE [9, 31]. On the one hand, this calls for future deep theoretical investigations of magnetic field triggered phonon side jump [42] and phonon skew scattering, which are indispensable for understanding room-temperature working principles of IPHE. On the other hand, the impurity skew scattering induced IPHE should be largely enhanced by increasing the mobility, suggesting a pathway to systematically engineering and improving the IPHE and its families in nonmagnetic materials.

In the scaling law analysis, we focused on the regime of $B \leq 3$ T, where the cubic scaling works perfectly. For stronger magnetic fields, the lowest-order perturbation breaks down and non-perturbative effects may become relevant, which introduce significant high-order contributions in magnetic field. Systematic understanding of the IPHE under such non-perturbative strong magnetic field [1, 4, 29] and the exploration of quantized IPHE in real materials are important open questions for future investigations.

**Methods**

**Single crystal growth.** High quality LuAuSn single crystals were grown by the Sn flux method. A mixture of pure elements with molar ratio Lu: Au: Sn = 1: 1: 10 was placed in an alumina crucible, which was then sealed in an evacuated quartz ampoule. The quartz ampoule was placed in a pit-type furnace and heated from room temperature to 1150 °C, where it was held for 24 hours. It was then slowly cooled to 650 °C at the rate of 2 °C/h. At this point, the excess Sn flux was removed by centrifugation.

**Single crystal characterization.** Single-crystal X-ray diffraction and crystal orientation were done at room temperature using a TD-3500 XRD system and TD-M3000 Laue system, respectively. The chemical composition of the LuAuSn sample was examined using scanning electron microscope (SEM) equipped with energy dispersive spectrometer (EDS). EDS measurements were performed at multiple positions on the crystal surfaces, with results consistent within the instrumental error margin of 1%−2%.

**Physical property characterization.** Magnetotransport properties were carried out using a Physical Property Measurement System (PPMS, 9 T). The angle-dependent electrical transport measurements were performed using the standard four-probe method with a horizontal rotator. A universal sample holder board was used for out-of-plane transport measurements, while a parallel magnetic field sample holder board was used for in-plane transport measurement.

**First-principles calculations.**

First-principles calculations based on density functional theory were performed using the Vienna ab initio Simulation Package [43-45] with the projector-augmented wave method [46]. The Perdew-Burke-Ernzerhof [47] generalized gradient approximation was employed to treat the exchange-correlation energy. The lattice constant was taken as 6.5627 Å. The inner positions are fully relaxed with force tolerance 0.01 eV/Å. The Γ-centered $k$-point mesh with size $8 \times 8 \times 8$ were used for Brillouin zone sampling. The Liechtenstein et al.'s approach [48] was used to treat localized f orbitals of Lu atoms with an effective $U = 6.7$ eV and $J = 0.7$ eV [34, 49]. The *ab initio* Wannier tight-binding models were constructed using the Wannier90 package [50]. A dense $k$-point mesh with size $300 \times 300 \times 300$ were used for Hall conductivity calculations. The intrinsic and the Lorentz force contributions were evaluated using first-principles calculations. The intrinsic response is given by $\sigma_{xy} = -(e^2/\hbar) \int f_0 \Omega_z$ [51,52], where $f_0$ is the Fermi-Dirac distribution function, and $\Omega_z$ is the Berry curvature. In calculations, both the band energy and Bloch states were dressed by considered the effect of magnetic field through the Zeeman's coupling $H_Z = \mu_B (\boldsymbol{L} + 2\boldsymbol{s}) \cdot \boldsymbol{B}/\hbar$, where $\mu_B$ is the Bohr magneton, $\boldsymbol{s}$ is the spin operator, and $\boldsymbol{L}$ is the orbital angular moment operator. The orbital angular momentum matrix elements in the basis of Bloch eigenstates are given by $\boldsymbol{L}_{mn} = m_e (\boldsymbol{\xi}_{ml} \times \boldsymbol{v}_{ln} - \boldsymbol{v}_{ml} \times \boldsymbol{\xi}_{ln})/2$ [53], where $\boldsymbol{\xi}_{mn} = (1-\delta_{mn})\boldsymbol{A}_{mn}$, $\boldsymbol{A}_{mn} = \langle u_m | i \partial_k u_n \rangle$ is the interband Berry connection, and $\boldsymbol{v}_{mn}$ is the matrix element of velocity operator. The in-plane Hall effect due to the Lorentz force was evaluated as $\sigma_{xy} = -(\tau e^3/\hbar) \int f_0' v^x \tau (\boldsymbol{v} \times \boldsymbol{B}) \cdot \partial_k v^y$ [54], where again, the band energy and Bloch states are dressed to include the Zeeman's coupling. The relaxation time was taken as $\tau = 0.1$ ps, estimated by comparing the measured longitudinal conductivity with the calculated Drude conductivity.

# Figures

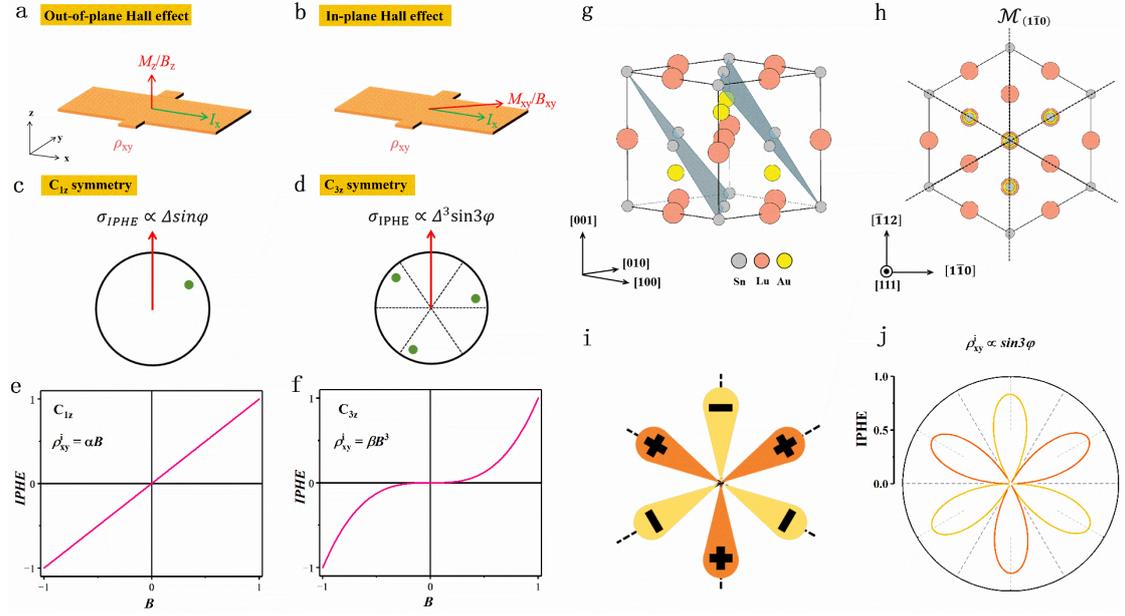

**Fig. 1 In-plane Hall effect and crystal structure of half-Heusler compound LuAuSn. a** and **b** schematic configurations out-of-plane and in-plane Hall effect measurement, respectively. **c** and **d** IPHE for $C_{1z}$ and $C_{3z}$ symmetries, respectively. **e** and **f** The expected *B*-linear and $B^3$-dependent behaviors for $C_{1z}$ and $C_{3z}$ symmetries, respectively. **g** The crystal structure of LuAuSn. The grey, orange and yellow spheres represent Sn, Lu and Au atoms, respectively. The grey planes are (111) planes composed of Sn atoms. **h** The top view of (111) plane and $C_{3z}$ symmetry. The three dashed lines represent mirror planes $\mathcal{M}_{(1\bar{1}0)}$. The $[\bar{1}12]$ and $[1\bar{1}0]$ directions are two non-equivalent axes in the (111) plane. **i** Schematic illustration of the $\varphi$-dependent IPHE and sign switch in the (111) plane. **j** The polar plot of IPHE in systems with $C_{3v}$ symmetry.

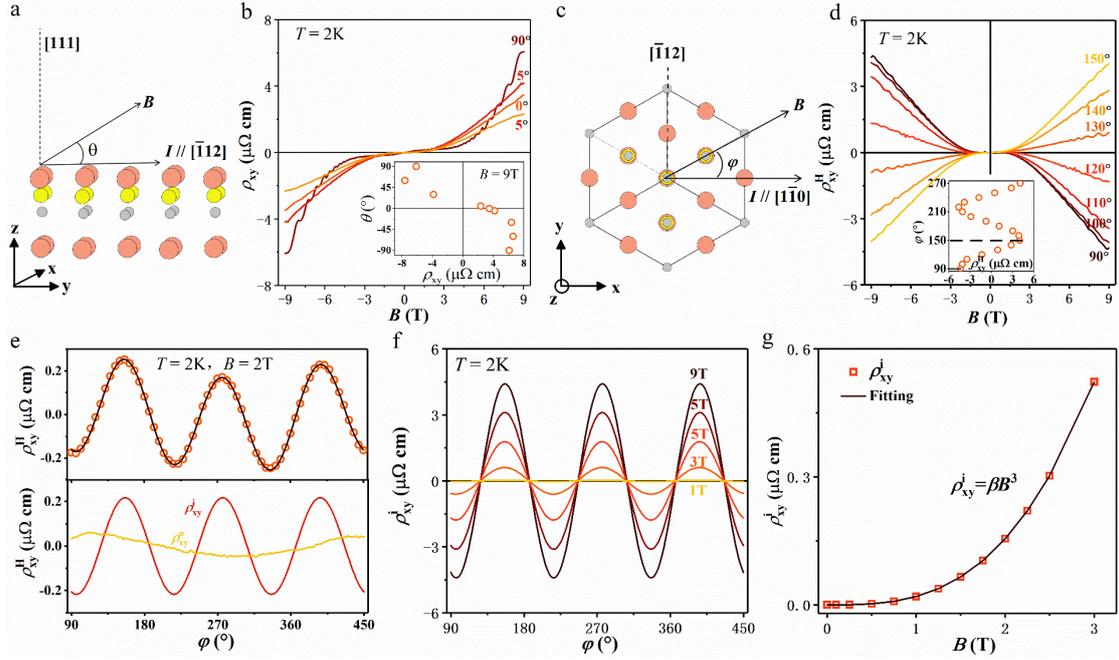

**Fig. 2 Observation of IPHE and $B^3$-dependent behavior for sample #1. a** The Hall effect measurement configuration for $B$ rotating from $[\bar{1}12]$ (in-plane, y axis) to $[111]$ (out-of-plane, z axis) direction. The xy plane corresponds to (111) plane. $\theta$ is the angle between $B$ and $[\bar{1}12]$ direction. **b** The Hall resistivity, $\rho_{xy}$, as a function of magnetic field measured at $\theta = -5°, 0°, 5°$ and $90°$ at $T = 2$K. The inset is $\theta$-dependent $\rho_{xy}$ at $B = 9$ T. **c** The measurement configuration for IPHE. $B$ rotates in (111) plane and current $I$ is applied along $[1\bar{1}0]$ directions. $\varphi$ is defined as the angle between $B$ and the $[1\bar{1}0]$ direction. **d** At $T = 2$K, the $B$-dependent Hall resistivity $\rho_{xy}^H$ at $\varphi = 90° -$ 150°. The inset is the extracted value of $\rho_{xy}^H$ at $B = 9$ T and $\varphi$ from $90°$ to $270°$. **e** Upper panel: $\varphi$-dependent $\rho_{xy}^H$ at $B = 2$ T. The red line is the fitting curve obtained by equation (2). Lower panel: decomposition of the total Hall signal into two sinusoidal components with periods $2\pi/3$ and $2\pi$, corresponding to in-plane $\rho_{xy}^i$ and out-of-plane $\rho_{xy}^o$ contributions. **f** The extracted $\rho_{xy}^i$ in different $B$ at $T = 2$ K. **g** The cubic-$B$ fitting of $\rho_{xy}^i \propto B^3$ in $B \leq 3\,T$.

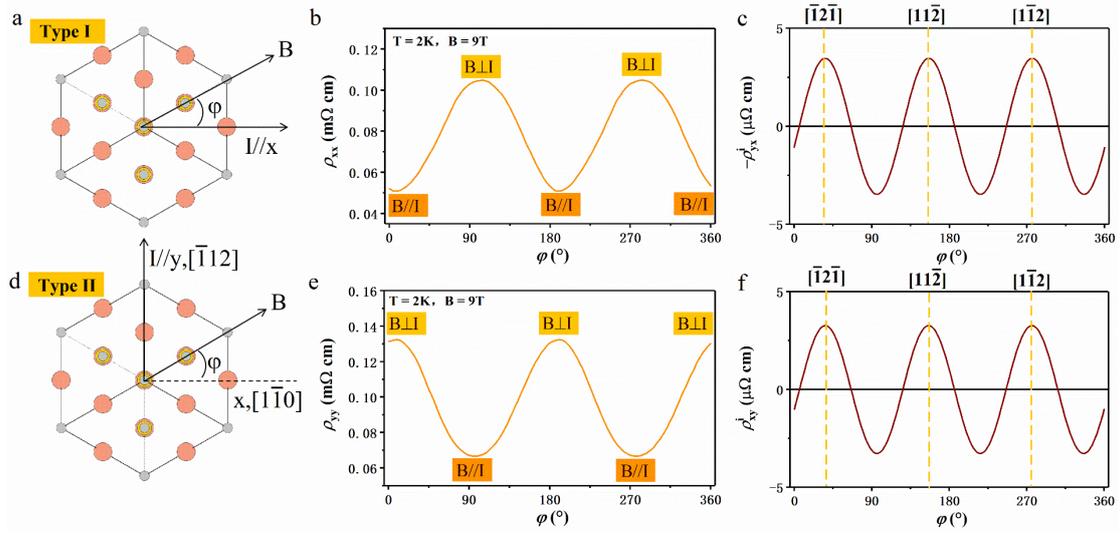

**Fig. 3 The in-plane anisotropic magnetoresistance (AMR) and IPHE with *I* // [1$\bar{1}$0] and [$\bar{1}$12] for sample #2. a** and **d** The schematic configurations of two measurement geometries Type I and Type II, respectively. For both types, *B* rotates in the (111) plane starting from [1$\bar{1}$0] direction. *φ* is defined as the angle between *B* and the [1$\bar{1}$0] direction, but *I* // [1$\bar{1}$0] for Type I and *I* // [$\bar{1}$12] for Type II. **b** and **e** The in-plane anisotropic magnetoresistance curves at 2K and 9T for Type I and II configurations, respectively. **c** and **f** Corresponding in-plane Hall resistivity curves.

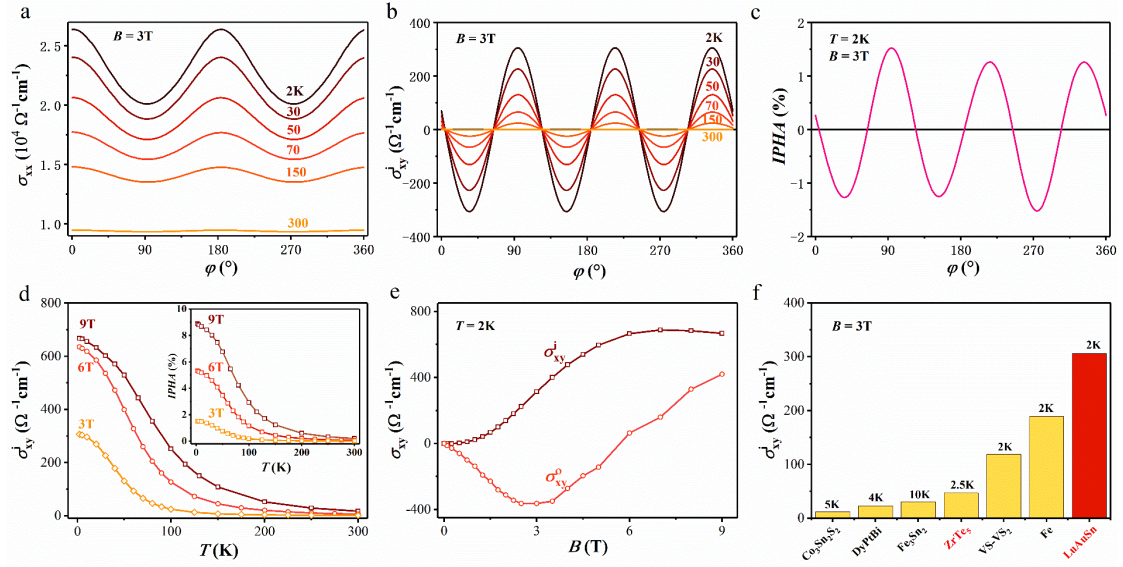

**Fig. 4 The in-plane Hall conductivity** $\sigma_{xy}^i$. **a** and **b** Angular dependent of longitudinal conductivity, $\sigma_{xx}$, and in-plane Hall conductivity, $\sigma_{xy}^i$, for different $T$ from 2 to 300 K at $B$ = 3 T, respectively. **c** The $\varphi$-dependent in-plane Hall angle (IPHA= $\sigma_{xy}^i(\varphi)/\sigma_{xx}(\varphi)$) at 2 K and 3 T. **d** The $T$-dependent of $\sigma_{xy}^i$ at $B$ = 9 T, 6 T and 3 T. The inset shows the temperature dependence IPHA. The peak value reaches to 8.9% at 2K and 9T. **e** The $B$-dependence of $\sigma_{xy}^i$ and out-of-plane Hall conductivity, $\sigma_{xy}^o$, at 2K. **f** The comparison of $\sigma_{xy}^i$ at 3T observed in different IPHE systems. ZrTe$_5$ and LuAuSn are nonmagnetic compounds. The data were taken from references [1-3, 5, 17, 18, 29].

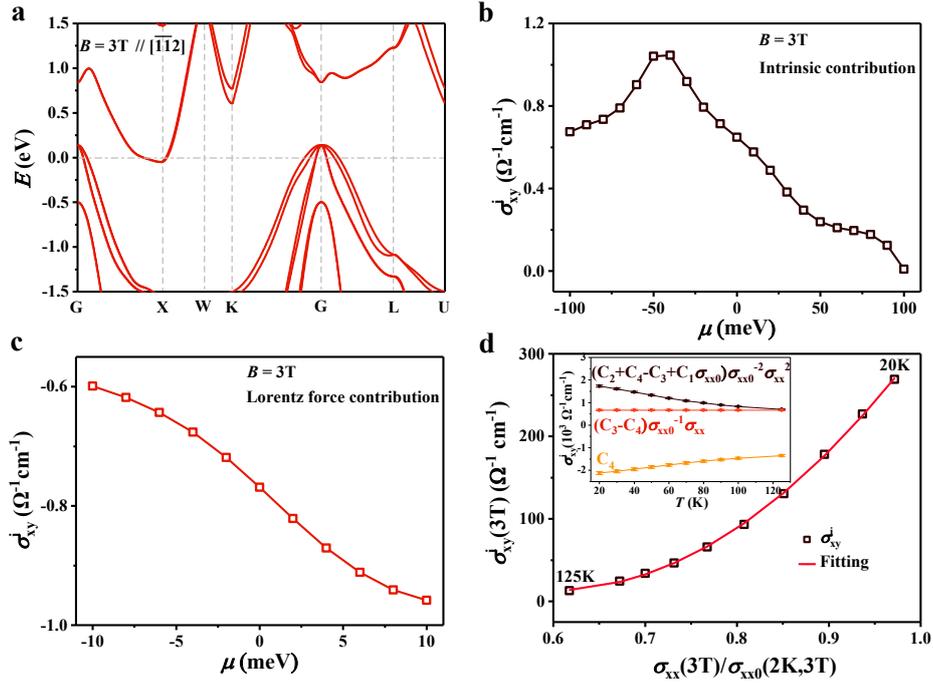

**Fig. 5 First-principles calculations and scaling analysis of the in-plane Hall conductivity in LuAuSn. a** The band structure of LuAuSn under an in-plane magnetic field of $B = 3\,T$ applied along $[\bar{1}\bar{1}2]$ direction. **b, c** Calculated intrinsic and Lorentz-force-induced in-plane Hall conductivity contributions. **d** The scaling of in-plane Hall conductivity $\sigma_{xy}^{i}(3T)$ as a function $\sigma_{xx}(3T)/\sigma_{xx0}(2K, 3T)$. Here, we take the in-plane longitudinal conductivity at 2K and 3T as the residual conductivity. The solid curve represents the fit using Eq. (2), and the individual contributing terms shown in the inset.